\documentstyle[12pt]{article}

\begin{document}

\baselineskip 0.7cm

\begin{titlepage}
\begin{flushright}
UT-03-17
\\
May, 2003
\end{flushright}

\vskip 1.35cm
\begin{center}
{\large \bf
Boundary Condition and the Cosmological Constant  
}
\vskip 1.2cm
S.~Hayakawa 
\vskip 0.4cm

{\it Department of Physics, University of Tokyo,\\
     Tokyo 113-0033, Japan}

\vskip 1.5cm

\abstract{
We discuss a model in which 
the boundary condition decides the four dimensional cosmological
 constant.
It is reviewed in a primitive way that boundary conditions are 
required by the action principle.
}

\end{center}
\end{titlepage}

\setcounter{page}{2}

\section{Introduction}

The extreme smallness of the cosmological constant relative to the
known scales is not easily explained by the field theoretical point
of view. 

There are four dimensional approaches such as
 \cite{Weinberg:1988cp},
 and higher dimensional approaches.
While six dimensional models are proposed    
\cite{Rubakov:1983bz},
five dimensional attempts are also made
\cite{Arkani-Hamed:2000eg},
\cite{Hayakawa:2001zf}.
Most of those approaches are to make the cosmological 
constant an integration constant, which doesn't depend on the Lagrangian
parameters. 
Higher dimensional approach has a benefit to use the freedom of the  
 extra dimension(s), but has constraints from regularity and         
necessity of compactification.                                       
With these difficulties, five dimensional models are expected to need
some fine tuning to vanish the four-dimensional cosmological constant
\cite{Forste:2000ps}.

We find the boundary condition is another possibility 
to make the cosmological constant vanish (or small)
without fine tuning of the Lagrangian parameters.






In this paper, we discuss a five dimensional model and consider its 
background configuration. Effective cosmological constant $\Lambda_4$
can be chosen zero by fine tuning of the boundary condition. 

\section{Boundary Condition}
According to the action principle, we have to put proper boundary 
conditions on fields in any field theory.
From now on, we think only Lagrangians 
which depend on the derivatives of matter fields only through the
bilinear terms of their first derivatives.
Then we get field equations of second derivative and boundary conditions
up to first derivative.
For example, consider the following Lagrangian with a boundary
(neglecting the gravity);
\begin{eqnarray}
 \int dx^{D} L =  
   \int dx^{D} {1 \over 2} \{\partial_{M}\phi \, 
     \partial^{M}\phi -m^2 \phi^{2} \}.
\end{eqnarray}
The action principle gives the equation of motion in the bulk,
as well as a constraint on the boundary;
\begin{eqnarray}
\int_{\partial} dx^{D} \partial^{M}\phi \, \delta \phi =0. \label{ex}
\end{eqnarray}
Above, $\delta \phi$ denotes the variation of $\phi$.
This constraint requires a boundary condition for $\phi$,
such as $\partial_{M}\phi=0$ or $\phi=0$ at the boundary.
A Lagrangian and a boundary condition together give one theory.

For any compact space-time with no boundary, 
which can be made from space-time with boundaries 
by identifying the corresponding points, 
the 
Lagrangian must be of equal value on identified points.
This means further constraint on boundary conditions.
Periodical condition is popular  
and this demands the values of the fields and their first derivatives to be
separately equal on identified points.
This actually satisfies the two constraints, 
vanishing of the boundary contribution in the variation  and the
equal values of the Lagrangian on boundaries.
For the four dimensional infinite Minkowski space-time, we can regard it as 
a compact box, put a periodical boundary condition and then make the 
size of the box infinite. Then we get plane wave solutions of free
particles.
Plane wave solutions are guaranteed by the symmetry of the 
Lagrangian and the background metric;
translational symmetry.
When we think a space-time without translational symmetry,
we need other continuous or
discrete symmetry for the Lagrangian and the metric
to identify the separate points.
Some of these symmetries are often used in identification, such as
twisted boundary condition under a global symmetry or gauge field
configuration with winding numbers.  


\section{The Model}

Here we restrict our attention to Lagrangians in which the fields 
 are already canonically normalized.

Let's think a model with $n$ massless real scalars and a 3-brane with
 negative tension $\lambda$ in a five dimensional bulk of positive 
cosmological constant $\Lambda$, which is 
 a generalization of the model \cite{Hayakawa:2001zf}. 
The action is
\begin{eqnarray}
 \int dx^{5} \sqrt{g} \{
 {1 \over 2}R -\Lambda + {1 \over 2}\displaystyle \sum_{i=0}^{n} 
\partial_{M}\phi_{i} \partial^{M}\phi_{i}  - \lambda \delta(y)  
\}, \label{action}
\end{eqnarray}
where $\Lambda > 0$ and $\lambda < 0$.
Note that here we set the five dimensional Planck scale $M_{5}=1$ 
for convenience.
We suppose a warped metric as a background;
\begin{eqnarray}
 ds^2 = \sigma(y)g_{\mu \nu}dx^{\mu}dx^{\nu} - dy^2, \label{warp}
\end{eqnarray}
where $g_{\mu \nu}$ denotes the four dimensional metric.
The Einstein equations and field equations for the background
configuration eq.(\ref{warp}) and $\phi(y)$ are
\begin{eqnarray}
  & & ~R_{\mu \nu}^{(4)} - {1 \over 2}g_{\mu \nu}R^{(4)}
          -{3 \over 2} g_{\mu \nu}\sigma'' = 
      \sigma g_{\mu \nu}\biggl( \Lambda + \displaystyle \sum_{i=1}^{n}   
             {1 \over 2}{{\phi_i}'}^2 \biggr)
       + \sigma g_{\mu \nu} \lambda \delta(y), \label{mn} \\
 & & -{3 \over 2}{{{\sigma'}^2} \over {\sigma}^2} 
       -{1 \over 2}\sigma^{-1}R^{(4)} =
        \Lambda - \displaystyle \sum_{i}{ 1 \over 2}{{\phi_i}'}^2 , \\
 & &  \phi_{i}'' + 2{{\sigma'} \over {\sigma}}\phi_{i}' = 0.
\end{eqnarray}
Here $R^{(4)}_{\mu \nu}$ and $R^{(4)}$ respectively mean the Ricci tensor and
scalar made of $g_{\mu \nu}$, and $~'$ denotes $\partial_{y}$.
The negative tension 3-brane causes a junction condition. That is,
integrating eq.(\ref{mn}) for $-\epsilon \le y \le +\epsilon$
gives 
\begin{eqnarray}
 {{\sigma'} \over {\sigma}} {\bigg|}_{-0}^{+0}= - { 2 \over 3}\lambda. 
\label{jc}
\end{eqnarray}

The Einstein equations in the bulk are reduced to
\begin{eqnarray}
 & & \Lambda_{4} - { 3 \over 2}\sigma '' 
      = \Lambda \sigma + \displaystyle \sum_{i=1}^{n}   
             {1 \over 2}\sigma {{\phi_i}'}^2, \\
 & & 2 {{\Lambda_4} \over {\sigma}} - { 3 \over 2} 
      {{{\sigma'}^2} \over {\sigma}^2} 
        = \Lambda - \displaystyle \sum_{i}{ 1 \over 2}{{\phi_i}'}^2 , \\ 
\end{eqnarray} 
where $\Lambda_4$ is the
four-dimensional cosmological constant and an
 integration constant of the bulk equations.
The above equations have an exact background solution for 
$\Lambda_4=0$. That is;
\begin{eqnarray}
 & & \phi'_{i}= a_{i} {\sigma}^{-2}, \label{sol1} \\
 & & \sigma^{2}(y) = \sqrt{\displaystyle \sum_{i} a^{2}_{i}} 
       \cos(\sqrt{{8\Lambda} \over 3}y + \alpha). \label{sol2} 
\end{eqnarray}
$a_{i}$'s and $\alpha$ are integration constants.
Since this has a parity symmetry, there is a possibility to
identify the points at $y=0$ and 
$y=y_{0} \equiv -2\alpha \sqrt{3 \over {8\Lambda}}$.
At least for the metric part, the values on both points are equal
and the discontinuity of the first derivative can be 
recognized as the result of the junction condition (\ref{jc}).
Then the junction condition (\ref{jc}) decides the value of $\alpha$,
and we get a geometry compactified along $y$ with no singularity.
The parity guarantees
the Lagrangian be equal on identified points;
the second constraint of compactification.
So far there need no fine tuning.

%
%
To vanish the boundary contribution in the variation,
proper boundary conditions for scalar fields are needed.
The variation $\phi_{i} + \delta \phi_{i}$ causes the following
term on the boundary;
\begin{eqnarray}
 \int dx^{4} \sqrt{g}\{ \displaystyle \sum_{i} 
   \partial^{y} \phi_{i} \delta \phi_{i}\}{\Big|}_{0}^{y_0}. \label{bd}
\end{eqnarray}
This time we already know a wanted configuration (\ref{sol1}),
so the problem is to seek for a boundary condition which makes eq.(\ref{bd})
vanish and which is satisfied by the solution (\ref{sol1}).

Let's give two examples.
One is for $n=1$,
\begin{eqnarray}
 {{\phi} \over {{M_{5}}^{3/2}}} {\bigg|}_{0}^{y_{0}} = c \label{bc1}
\end{eqnarray}
where
 $c$ is a given constant.
The other example is for $n=2$, the boundary condition of the following 
non-linear form;
\begin{eqnarray}
 A {{\phi_{1}(0)}\over {\phi_{1}(y_0)} } 
  + B {{\phi_{2}(0)} \over {\phi_{2}(y_0)}}
   + C {{\phi_{1}(0)}\over {\phi_{2}(0)}} + D =0, \label{bc2}
\end{eqnarray}
where $A,B,C,D$ are constants.
In both cases, we also put the condition
\begin{eqnarray}
 \partial_y \phi_{i}|_{\partial}=0, \label{bc3}
\end{eqnarray}
which is satisfied by eq.(\ref{sol1}).
With one fine tuning of $c$ or $\{ A,B,C,D \}$, each boundary
condition 
make eq.(\ref{bd}) vanish,
and also includes the solution (\ref{sol1}).
The variation of eq.(\ref{bc1}) is $\delta \phi(0) = \delta \phi(y_{0})$,
and it is clear that this and the solution(\ref{sol1}) make
eq.(\ref{bd}) vanish. The fine tuning is
\begin{eqnarray}
 c = \int_{0}^{y_0}\cos^{-1}(\sqrt{{8\Lambda} \over 3}y + \alpha).
\end{eqnarray}
For the boundary condition (\ref{bc2}), the variation is
\begin{eqnarray}
 & & \biggl( {A \over {\phi_1(y_0)}} + {C \over {\phi_2(0)}} \biggr) 
   \delta \phi_{1}(0) 
  - A {{\phi_1(0)} \over {{\phi_1(y_0)}^2}} \delta \phi_1(y_0) \nonumber \\
 & & ~~~~~~~
 +\biggl( {B \over {\phi_2(y_0)}} - C{{\phi_1(0)} \over {{\phi_2(0)}^2}} 
    \biggr) \delta \phi_2(0)
   - B {{\phi_2(0)} \over {{\phi_2(y_0)}^2}} \delta \phi_2(y_0) =0.
\end{eqnarray}
To vanish eq.(\ref{bd}) we have to put
\begin{eqnarray}
 & & {A \over {\phi_1(y_0)}} + {C \over {\phi_2(0)}}
      =  A {{\phi_1(0)} \over {{\phi_1(y_0)}^2}} \\
 & & {B \over {\phi_2(y_0)}} - C{{\phi_1(0)} \over {{\phi_2(0)}^2}} 
      = B {{\phi_2(0)} \over {{\phi_2(y_0)}^2}} \\
 & & a_2 A {{\phi_1(0)} \over {{\phi_1(y_0)}^2}} 
      = a_1 B {{\phi_2(0)} \over {{\phi_2(y_0)}^2}}. 
\end{eqnarray}
And there is the forth condition for the solution (\ref{sol1}) to 
satisfy eq.(\ref{bc2}).
We have three free parameters, $\phi_1(0),\phi_2(0)$ and ${a_1}/{a_2}$,
so there need one fine tuning between $A,B,C,D$.

These boundary conditions are invariant under the scale transformation
\begin{eqnarray}
\left\{
 \begin{array}{ll}
   \tilde{\phi}_{i} = k^{-3/2}\phi_{i}, \\
   \tilde{y} = k y, \\
   \tilde{x}^{\mu} = k x^{\mu},  \\
   \tilde{M}_{5} = k^{-1} M_{5}. 
 \end{array}
\right. \label{sc}
\end{eqnarray}
That is, the parameters $c$ or $A,B,C,D$ can not be transformed 
by (\ref{sc}).
For the proper choice of $c$ or $\{ A,B,C,D \}$, the boundary conditions
(\ref{bc1}), (\ref{bc2}) and (\ref{bc3}) %
designate a $\Lambda_{4}=0$ solution, independent of 
values of the Lagrangian parameters. 

On the other hand, if the parameters in a boundary condition is variant under
the above transformation (\ref{sc}), the fine tuning of the boundary condition 
can be transformed to the fine tuning of the five-dimensional 
Planck scale $M_{5}$.
One choice of the parameters of the boundary condition leads to
different values of $\Lambda_4$ depending on the value of $M_{5}$.


\section{Conclusion}



We have proposed a new approach to the cosmological constant problem;
fine tuning by the boundary condition.
Since the basic purpose is to make the four-dimensional cosmological 
constant independent of the known scales,
the boundary conditions 
eq.(\ref{bc1}), eq.(\ref{bc2}) and eq.(\ref{bc3}) must not have
other scale parameters than Lagrangian parameters. 
This is the another reason why the boundary conditions 
need to be invariant under the scale transformation (\ref{sc}). 


Note that here we have used a constant shift symmetry of $\phi_{i}$, which is 
peculiar to the model (\ref{action}).

Let's consider more about the model (\ref{bc1}).
The non-zero $\partial_y \phi$ in the background solution only 
contributes to the cosmological constant through the energy-momentum
tensor. Since this is a zero-mode solution, the ``momentum'' 
$\partial_y \phi$ doesn't affect the energy of $\phi$.
Also the excitation has usual periodical boundary condition and don't
know whether its background configuration is periodical or not.   

In this paper we have primitively taken compactification to be defined 
by the values of the fields and their first derivatives.
Much familiar way is to relate $\phi(y)$ and $\phi(y+y_0)$;
when $\phi$ is a function of $C^{\infty}$ it ensures any higher
derivative to be related at identified points.
Actually in the case of (\ref{bc1}), we can rewrite the condition to
\begin{eqnarray}
 \phi(y+y_0) = \phi(y) +c
\end{eqnarray}
without any change of the theory.

Although we have proposed a five dimensional model, somehow there might be 
a four dimensional theory in which the cosmological constant is
also decided by the boundary condition
considering some boundaries such as black holes or initial conditions.


\vspace{10mm}

\noindent
{\Large\bf Acknowledgments}

We would like to thank T.~Yanagida and Izawa K.-I. for many useful 
discussions.

\end{document}